\documentclass[a4paper,11pt]{article}
\pdfoutput=1 
\usepackage{jinstpub} 
\usepackage[utf8]{inputenc}
\usepackage{textgreek}
\usepackage{subcaption}
\usepackage{lineno}
\usepackage{siunitx}
\usepackage{xcolor}

\title{\boldmath Impact of front-end parameters of the ARCADIA MD3 on charged particle detection}

\author[a,b,1]{C. Pantouvakis\note{Corresponding author.}}
\author[l]{S. Garbolino}
\author[a,b]{M. Rignanese}
\author[g]{P. Affleck}
\author[f]{A. Apresyan}
\author[b]{P. Azzi}
\author[b,f]{N. Bacchetta}
\author[b,c]{C. Bonini}
\author[a, b, e]{D. Chiappara}
\author[b,c]{S. Ciarlantini}
\author[m]{D. Falchieri}
\author[i]{A. Hayrapetyan}
\author[a,b]{S. Mattiazzo}
\author[n]{L. Pancheri}
\author[a,b]{D. Pantano}
\author[l]{A. Rivetti}
\author[l]{M. Rolo}
\author[o]{R. Santoro}
\author[b]{R. Turrisi}
\author[b,d]{J. Wyss}
\author[f,h]{S. Xie}
\author[a,b]{A. Zingaretti}
\author[f]{I. Zoi}
\author[a,b]{P. Giubilato}

\affiliation[a]{Department of Physics and Astronomy "G. Galilei", University of Padova, Padova, Italy}
\affiliation[b]{INFN Padova, Italy}
\affiliation[c]{Centro di Ateneo di Studi e Attività Spaziali CISAS "G. Colombo", University of Padova, Padova, Italy}
\affiliation[d]{University of Cassino and Southern Lazio, DICEM, Cassino, Italy}
\affiliation[e]{CERN, Geneva, Switzerland}
\affiliation[f]{Fermi National Accelerator Laboratory, Batavia, IL, USA}
\affiliation[g]{University of Florida, Gainesville, Florida, USA}
\affiliation[h]{California Institute of Technology, Pasadena, CA, USA}
\affiliation[i]{A. I. Alikhanyan National Science Laboratory (Yerevan Physics Institute), Yerevan, Armenia}
\affiliation[l]{INFN Torino, Italy}
\affiliation[m]{INFN Bologna, Italy}
\affiliation[n]{INFN TIFPA, and Department of Industrial Engineering, University of Trento, Trento, Italy}
\affiliation[o]{INFN, Milano and Università dell’Insubria, Varese, Italy}

\emailAdd{caterina.pantouvakis@phd.unipd.it}

\abstract{
    The ARCADIA INFN R\&D project developed a Fully Depleted Monolithic Active Pixel Sensor (FD-MAPS) using a customized LFoundry \SI{110}{\nano\meter} CIS process. The first in-beam characterization of the ARCADIA Main Demonstrator 3 (MD3) sensor with \SI{200}{\micro\meter} active thickness has been performed at the Fermilab Test Beam Facility with a \SI{120}{\giga\eV} proton beam. The Device Under Test (DUT) is tested with a trigger-less telescope composed of two ARCADIA MD3 tracking planes. This early study investigates the effect of the front-end parameters on the tracking performance.
}


\proceeding{Topical Workshop on Electronics for Particle Physics\\
Rethymno, Aquila Rithymna Beach, Crete, Greece\\
06-10 October 2025}

\begin{document}
\maketitle
\flushbottom

\section{The ARCADIA MD3 Fully Depleted MAPS}
\label{sec:intro}

\begin{figure}[ht]
  \centering
  \begin{subfigure}[b]{0.49\textwidth}
    \includegraphics[height=4cm]{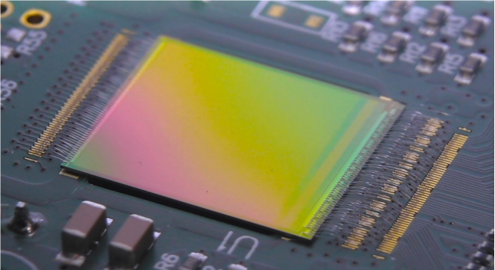}
    \caption{}
    \label{arcadia_chip}
  \end{subfigure}
  \begin{subfigure}[b]{0.49\textwidth}
    \includegraphics[height=4cm]{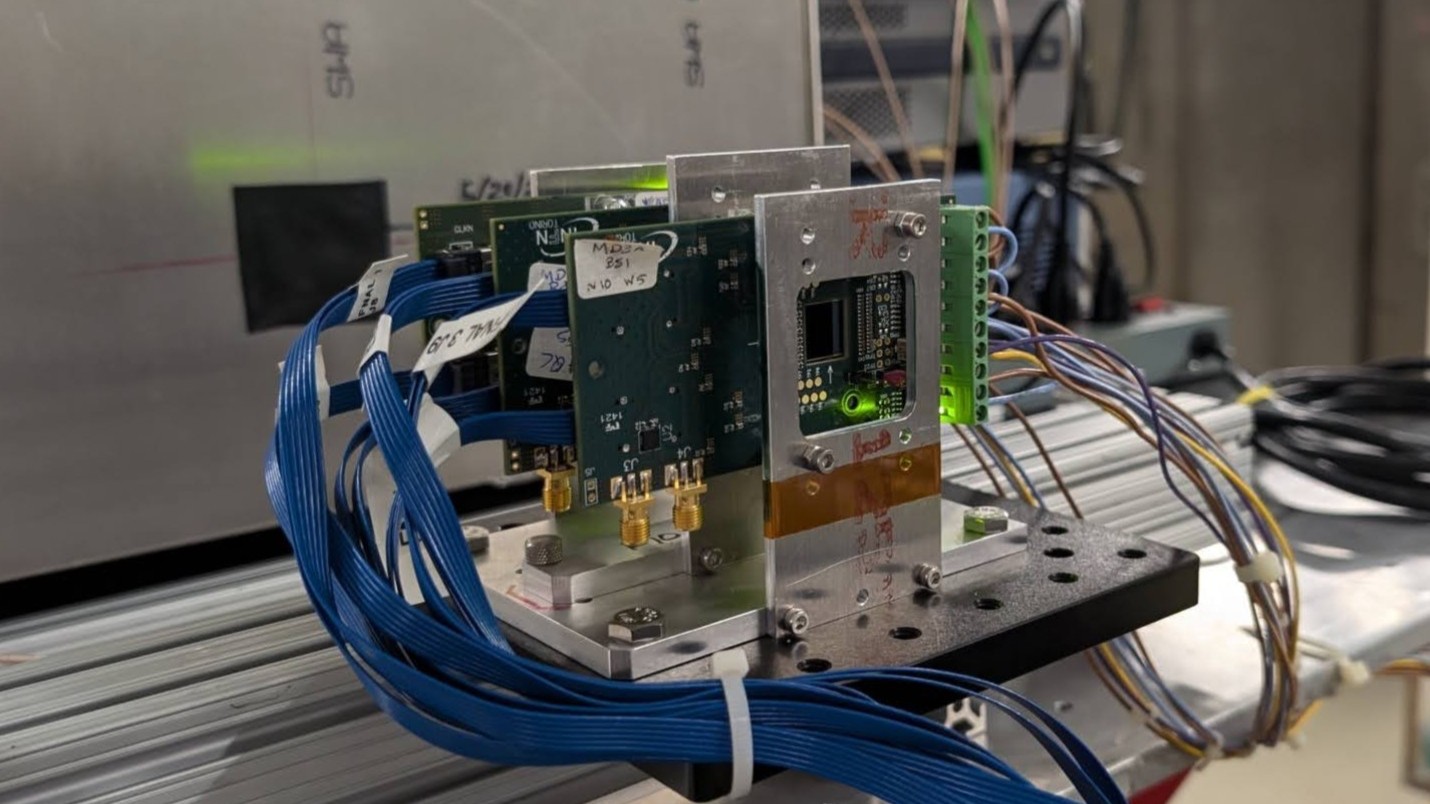}
    \caption{}
    \label{fig:test_beam_setup}
  \end{subfigure}
  \caption{\textcolor{black}{The} ARCADIA MD3 chip bonded on the PCB \cite{10.3389/fsens.2025.1603755} \subref{arcadia_chip} and \textcolor{black}{the} test beam setup \textcolor{black}{consisting of two MD3 tracking planes and one MD3 as DUT} \subref{fig:test_beam_setup}}
\end{figure}

The ARCADIA INFN R\&D project developed a thick, Fully Depleted Monolithic Active Pixel Sensor (FD-MAPS) technology demonstrator using a customized LFoundry 110nm CIS process. \\
\textcolor{black}{With respect to standard epitaxial MAPS, e.g. the ALPIDE chip \cite{MAGER2016434}, or the most recent prototypes developed for the ALICE ITS3 upgrade \cite{AGLIERIRINELLA_APTS}, the ARCADIA FD-MAPS features charge collection exclusively by drift, enhancing collection efficiency and radiation hardness, in fully depleted substrates with thicknesses of few hundred micrometers. This is achieved by employing} an n-type high resistivity substrate with an additional n-epitaxial layer to control the potential beneath the deep p-well hosting the CMOS circuitry. The whole high-resistivity substrate achieves full depletion with a uniform electric field, by applying a negative voltage to the backside p$^+$ implant, realized with a custom process \cite{9075426}.\\
\noindent The latest full chip demonstrator, called Main Demonstrator 3 (MD3, shown in \autoref{arcadia_chip}), sports a $512\times512$ pixel array, with a \SI{25}{\micro\meter} pixel pitch, for a total active area of $1.28\times1.28$ \SI{}{\square\centi\meter}. The digital pixel output feeds an event-driven readout architecture capable of handling rates up to \SI{100}{\mega\hertz\per\centi\meter\squared}. The chip design \textcolor{black}{is} optimized for very low power consumption (\SIrange{10}{30}{\milli\watt\per\centi\meter\squared}, depending on the event rate). The MD3 chip is available with active thicknesses of 50, 100, or 200 \si{\micro\meter}. The features of the chip, in particular the small pixel pitch and low power consumption, make it well suited for tracker detectors in future collider experiments, such as the FCC-ee (Future Circular Collider) as well as for space and medical applications. \textcolor{black}{In the context of tracker detectors at future collider experiments, a strong effort on the optimization of the chip, in all its design aspects, is needed to meet the detector requirements, which are becoming increasingly challenging. This study focuses on the impact of front-end bias currents on tracking performance.}

\subsection{Chip layout and readout}

\begin{figure}[ht]
  \centering
    \begin{subfigure}{0.49\textwidth}
    \includegraphics[height=5.5cm]{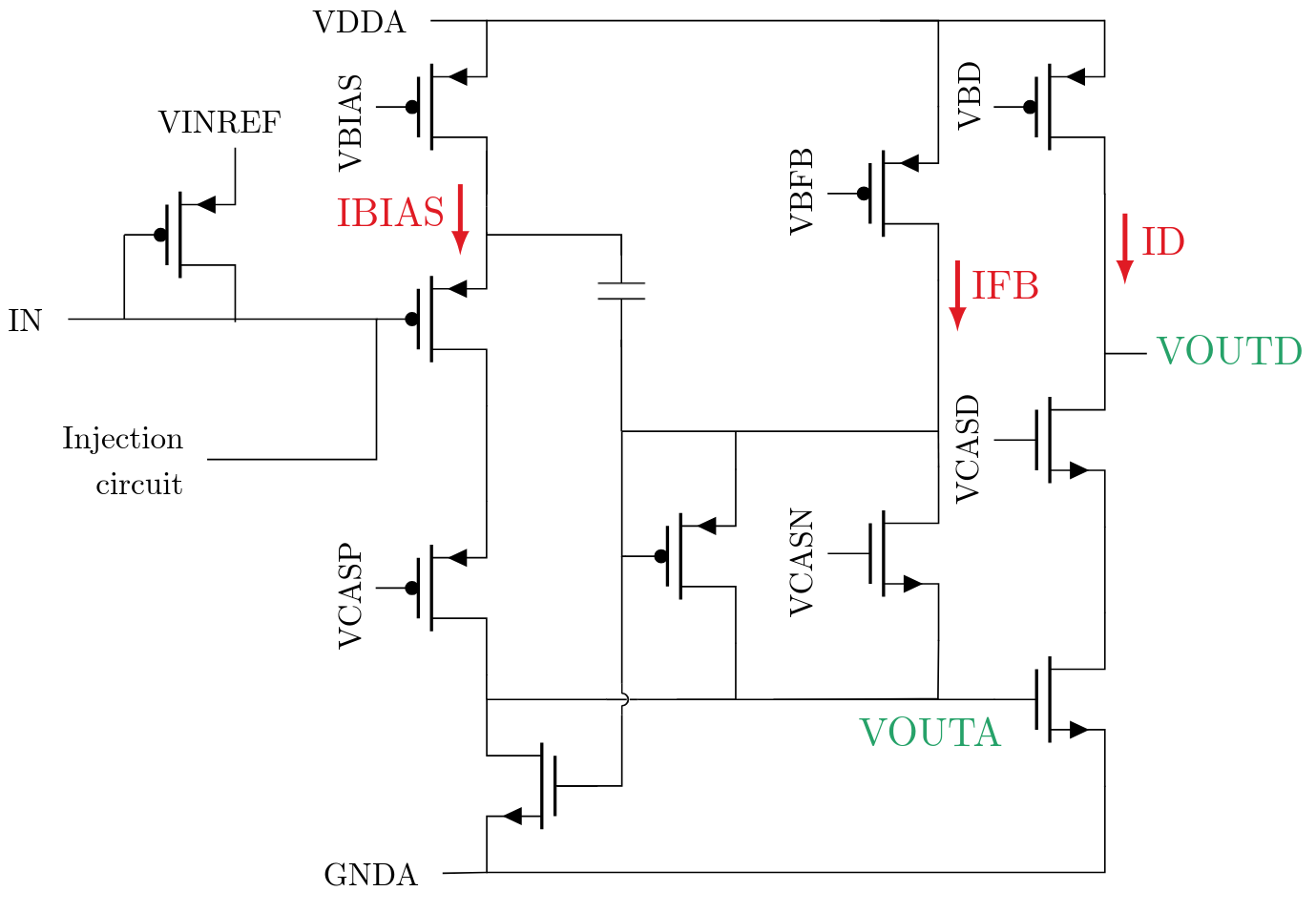}
    \caption{}
    \label{fig:fe_circuit}
    \end{subfigure}
    \begin{subfigure}{0.49\textwidth}
      \centering
    \includegraphics[height=6.25cm]{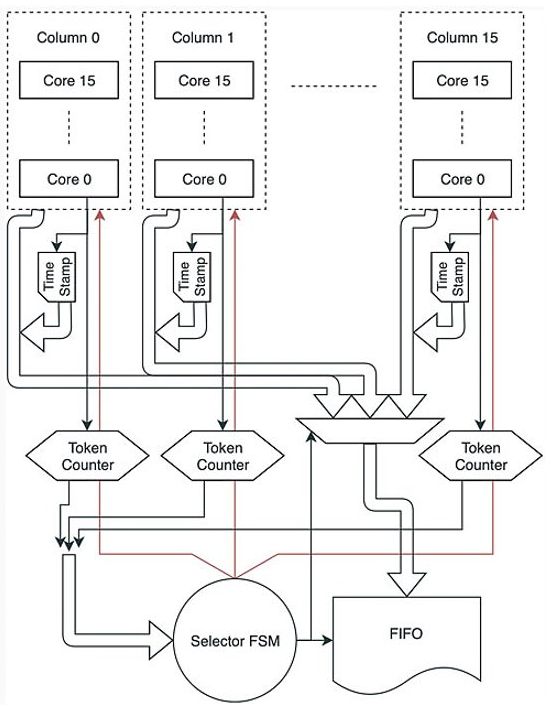}
    \caption{}
    \label{fig:arcadia_section_readout}
    \end{subfigure}
\caption{\textcolor{black}{Scheme of the analog front-end circuit \subref{fig:fe_circuit} and diagram of the section readout unit \cite{10.3389/fsens.2025.1603755}} \subref{fig:arcadia_section_readout}}
\label{fig:arcadia_diagrams}
\end{figure}

The pixel array is organized in 16 sections of $512\times32$ \textcolor{black}{pixels}, each with independent biasing and readout circuitry at the periphery. Pixels within the section are further organized in 2-pixel wide columns, and every column is segmented in 32 units, the so-called cores. The pixel core, which represents the minimum synthesizable unit, contains the pixel regions with the analog circuitry on the sides, the input buffers and output logic at the bottom. \\
\textcolor{black}{The in-pixel analog front-end circuit is an ALPIDE-like hit discriminator, details can be found in \cite{Kim_2016}. Being optimized for high gain, it is inherently non-linear and saturates quickly. The charge collected by the pixel diode generates a signal at the input of the front-end (see \autoref{fig:fe_circuit}), constituted by a high-gain amplifier. It consists of a main branch with a source-follower input transistor that copies the input voltage on a big capacitor connected to its source. The charge transfer to the output node is responsible for voltage amplification. The feedback branch sets the amplifier baseline. The output stage is a discriminator implemented as a simple common-source stage. A clipping mechanism for large input signals is also integrated. If the amplifier output is above threshold, an active low discriminator output is generated and the pixel status register is set to 1.} \\
The chip features a frame-less and clock-less asynchronous readout architecture to achieve the desired power consumption figures. \textcolor{black}{As schematized in \autoref{fig:arcadia_section_readout},} whenever a pixel receives the writing token, the hitmap, region and core addresses are propagated to the section readout unit, and a timestamp is latched. The payload, consisting of the column data, column address, and timestamp, is then sent to the output FIFO. The 32-bit data words are 8b10b encoded in 40-bit packets and sent out via 16, one per section, 100-320 MHz DDR seriali\textcolor{black}{z}ers.

\section{Test beam experimental setup}
\textcolor{black}{The} test beam measurements were performed with a 120 GeV proton beam at the Fermilab Test Beam Facility (FTBF) in Summer 2024.
This facility provides the beam with \textcolor{black}{a} spill repetition rate of 1 minute, and each spill lasts \SI{4.2}{s}. Each spill consists of streams of 7 batches coming from the Main Injector, where each batch is \SI{11.2}{\micro\second} long \cite{fermilab}. \\
The experimental setup, shown in \autoref{fig:test_beam_setup}, consists of two ARCADIA MD3 as tracking planes (planes 0 and 2) and one ARCADIA MD3 as Device Under Test (DUT), all with an active thickness of \SI{200}{\micro\meter}. The backside bias of the three planes \textcolor{black}{was} set to \textcolor{black}{\SI{-80}{\volt}}, \SI{-90}{\volt}, \SI{-90}{\volt} to have full depletion. Assuming the DUT z position at $z_{DUT} = 0$, the two tracking planes \textcolor{black}{were} located at $z_0 = \SI{-33.8}{\milli\meter}$ and $z_2 = \SI{+29.4}{\milli\meter}$ with respect to the DUT. The DUT was placed perpendicularly to the beam to evaluate the tracking performance with orthogonal charged particle tracks.

\section{Data analysis and results}
Several measurements \textcolor{black}{were} performed in the test beam campaign to characterize the tracking performance of the ARCADIA MD3 DUT. \\
The DAQ system \textcolor{black}{relied} on trigger-less acquisition with timestamp synchronization of the three planes.
Due to the structure of the beam, \textcolor{black}{the} data \textcolor{black}{were} separated \textcolor{black}{into} spills by looking at the distribution of \textcolor{black}{the} packet timestamps. \\
Data analysis \textcolor{black}{was}
performed spill-by-spill using Corryvreckan software for test beam studies \cite{Dannheim_2021}.

\begin{figure}[ht]
  \centering
    \includegraphics[width=8cm]{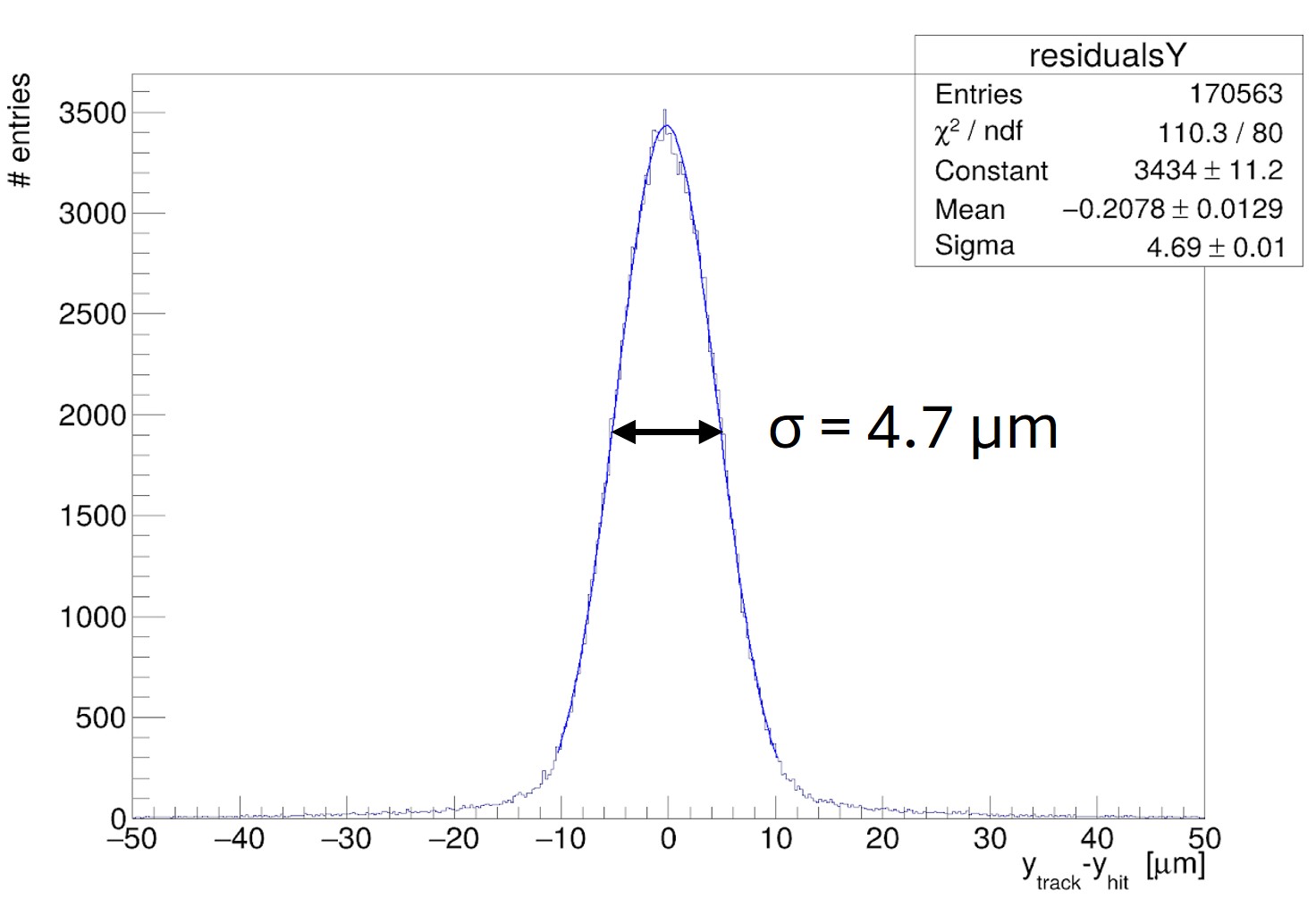}
    \caption{Example of \textcolor{black}{the distribution of the residuals in the y-dimension of the DUT}}
    \label{fig:gaussian fit}
\end{figure}

\noindent The first analysis steps include \textcolor{black}{hit} clusterization and \textcolor{black}{the} pre-alignment of the DUT and the second external plane with respect to the first tracking plane, which acts as \textcolor{black}{the} reference plane, using spatial correlation plots. After pre-alignment, the \textcolor{black}{positions} of the tracking planes \textcolor{black}{are} fixed, and the DUT is corrected for translational and rotational misalignment, considering the residual distributions. To estimate \textcolor{black}{the} spatial resolution and efficiency, tracks are reconstructed with the two external planes, and clusters on the DUT are associated \textcolor{black}{with} the tracks with an association window of \SI{5}{\micro\second} and \SI{500}{\micro\meter} and an edge cut of 30 pixels on both rows and columns. An example of the distribution of the DUT residuals along the row axis is shown in \autoref{fig:gaussian fit}.

\subsection{Front-end parameters scan}
\textcolor{black}{The front-end bias voltages and currents are adjustable via digital-to-analog converters (DACs), and the fine tuning of the front-end parameters is essential to characterize and optimize the pixel response.} The main parameter that controls the pixel threshold is the VCASN \textcolor{black}{bias voltage, which sets the amplifier output baseline and is} connected to a 6-bit DAC. \textcolor{black}{In addition, there are two front-end bias currents that modify the effective pixel threshold: the discriminator current ID and the bias current in the feedback branch IFB, controlled by \textcolor{black}{a} 2-bit DAC each.} 
\textcolor{black}{The discriminator current ID is directly proportional to the threshold and it also influences the slope of the discriminator output. The IFB current acts on the amplifier output signal, affecting both the baseline level and the speed at which it returns to the baseline. An increase of IFB leads to a decrease of the baseline, leading to an effective increase of the pixel threshold. By design, this current should be kept equal to the bias current in the main branch, IBIAS, to optimize the amplifier operating point and speed.} \\
\textcolor{black}{Scans of these front-end parameters were performed to evaluate the effect on the DUT tracking capabilities.} While scanning these settings, the VCASN parameter \textcolor{black}{was} kept fixed at 5, the chosen reference value, which corresponds to a much lower charge with respect to the signal of a Minimum Ionizing Particle. In the next subsection, the results of the scan on ID and \textcolor{black}{the} scan on both IBIAS and IFB are presented. 
\subsection{Results}
\begin{figure}[ht]
 \centering
  \begin{subfigure}[b]{0.49\textwidth}
    \includegraphics[width=7.25cm]{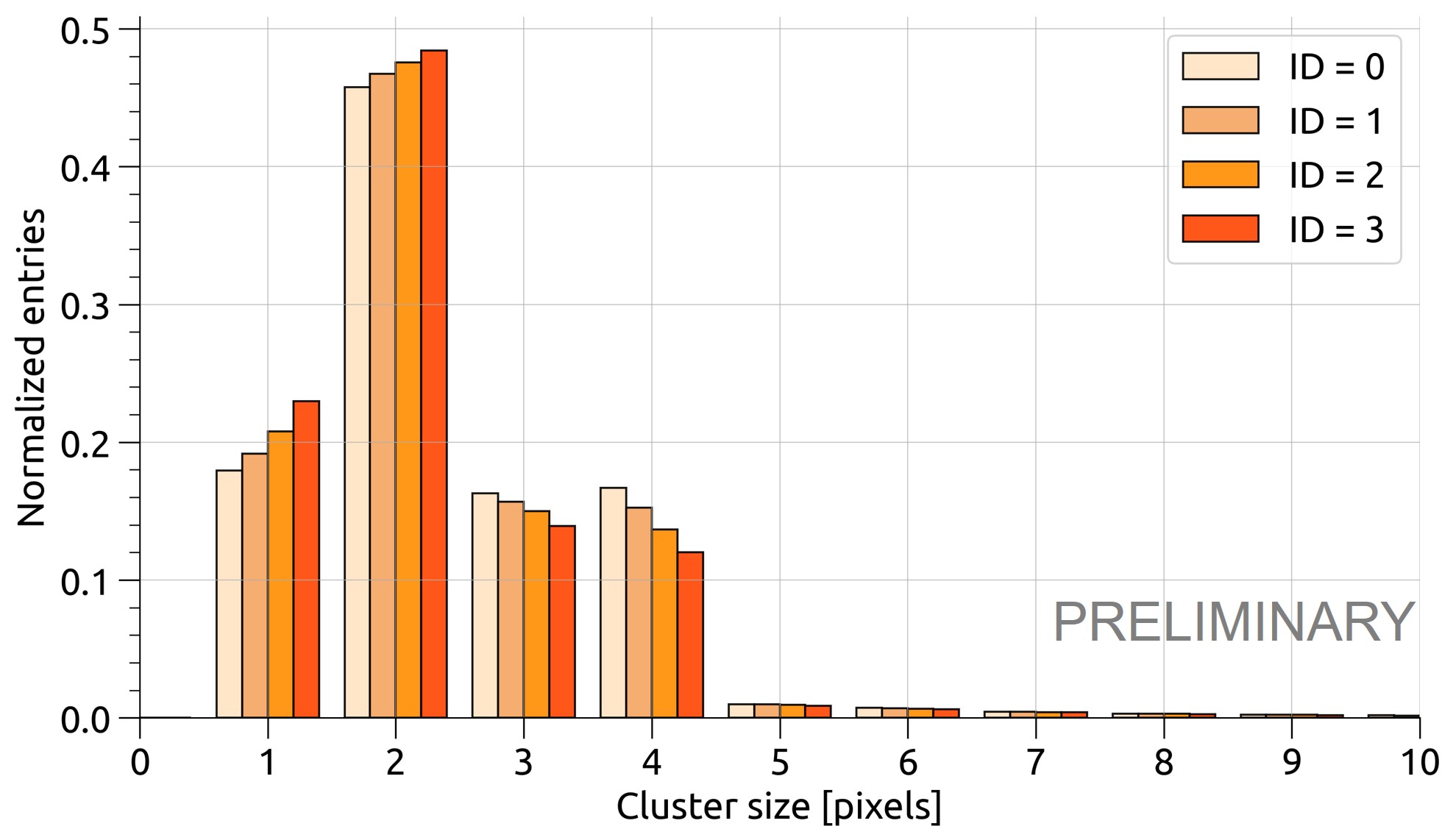}
    \caption{}
    \label{fig:clz_id}
  \end{subfigure}
  \begin{subfigure}[b]{0.49\textwidth}
    \includegraphics[width=7.25cm]{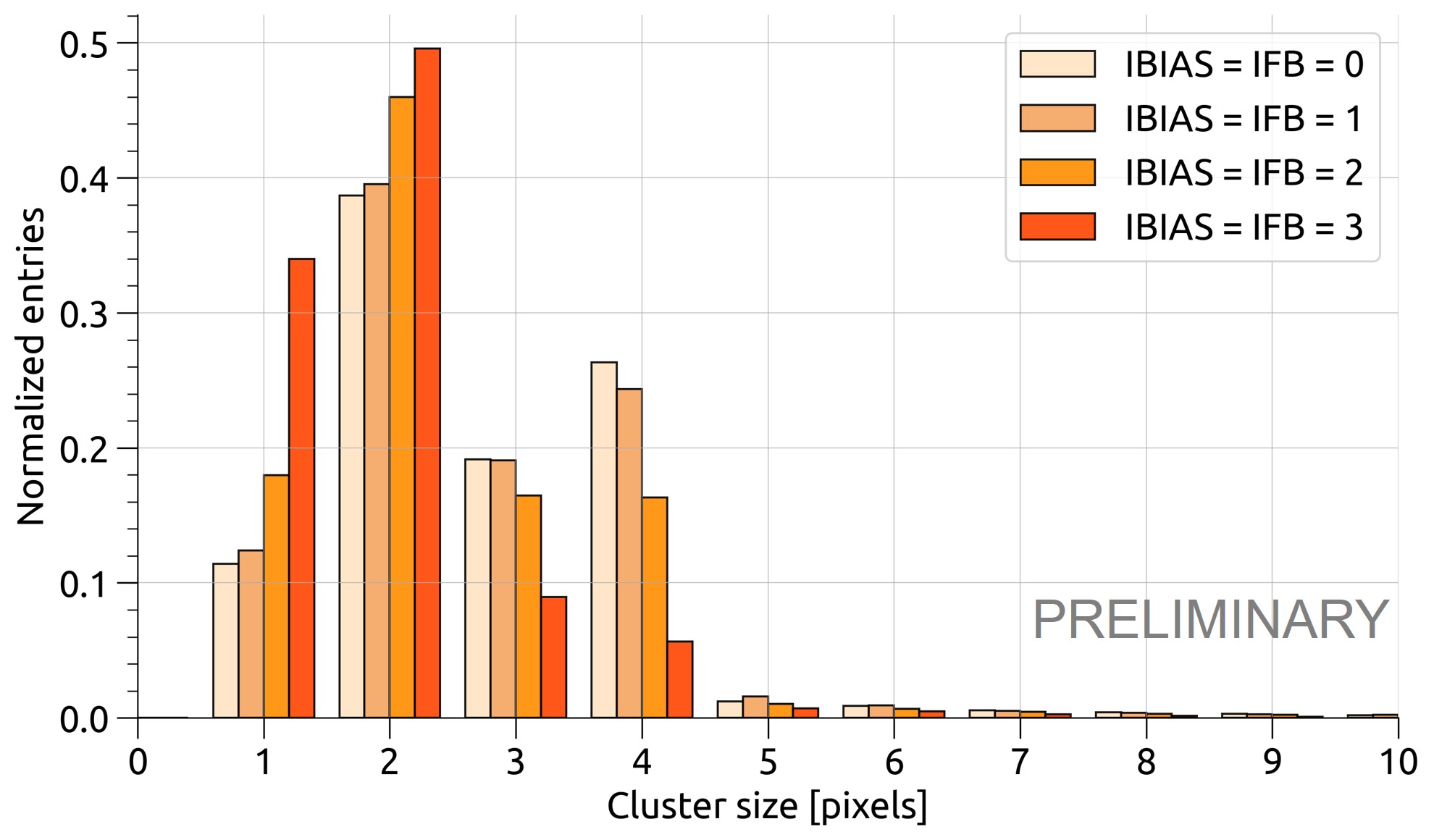}
    \caption{}
    \label{fig:clz_ibias}
  \end{subfigure}
  \caption{Cluster size distribution for ID \subref{fig:clz_id} and IBIAS-IFB \textcolor{black}{scan} \subref{fig:clz_ibias}}
\end{figure}

\begin{figure}[ht]
  \centering
  \begin{subfigure}[b]{0.49\textwidth}
      \centering
    \includegraphics[width=7.25cm]{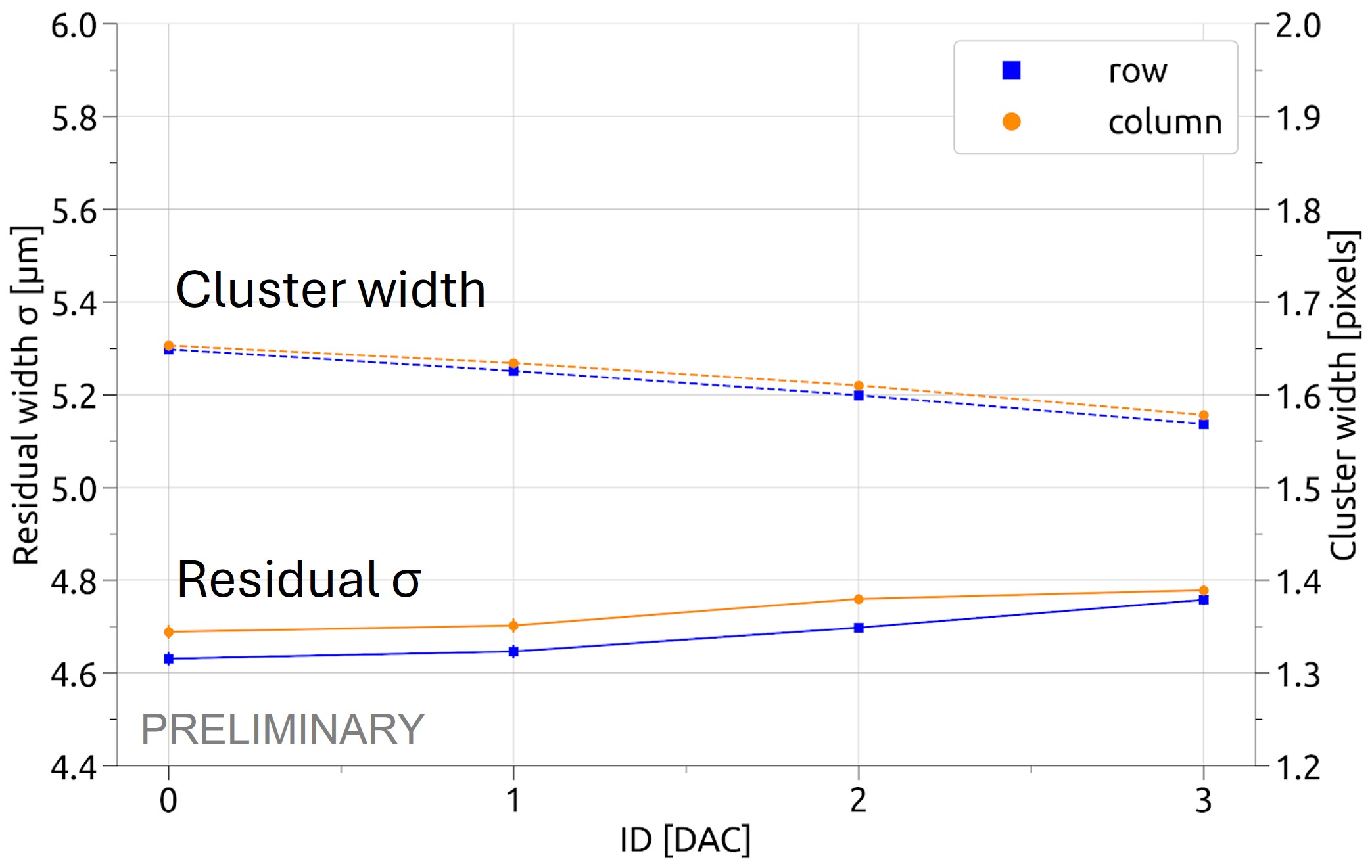}
    \caption{}
    \label{fig:id_resolution}
  \end{subfigure}
  \begin{subfigure}[b]{0.49\textwidth}
    \centering
    \includegraphics[width=7.25cm]{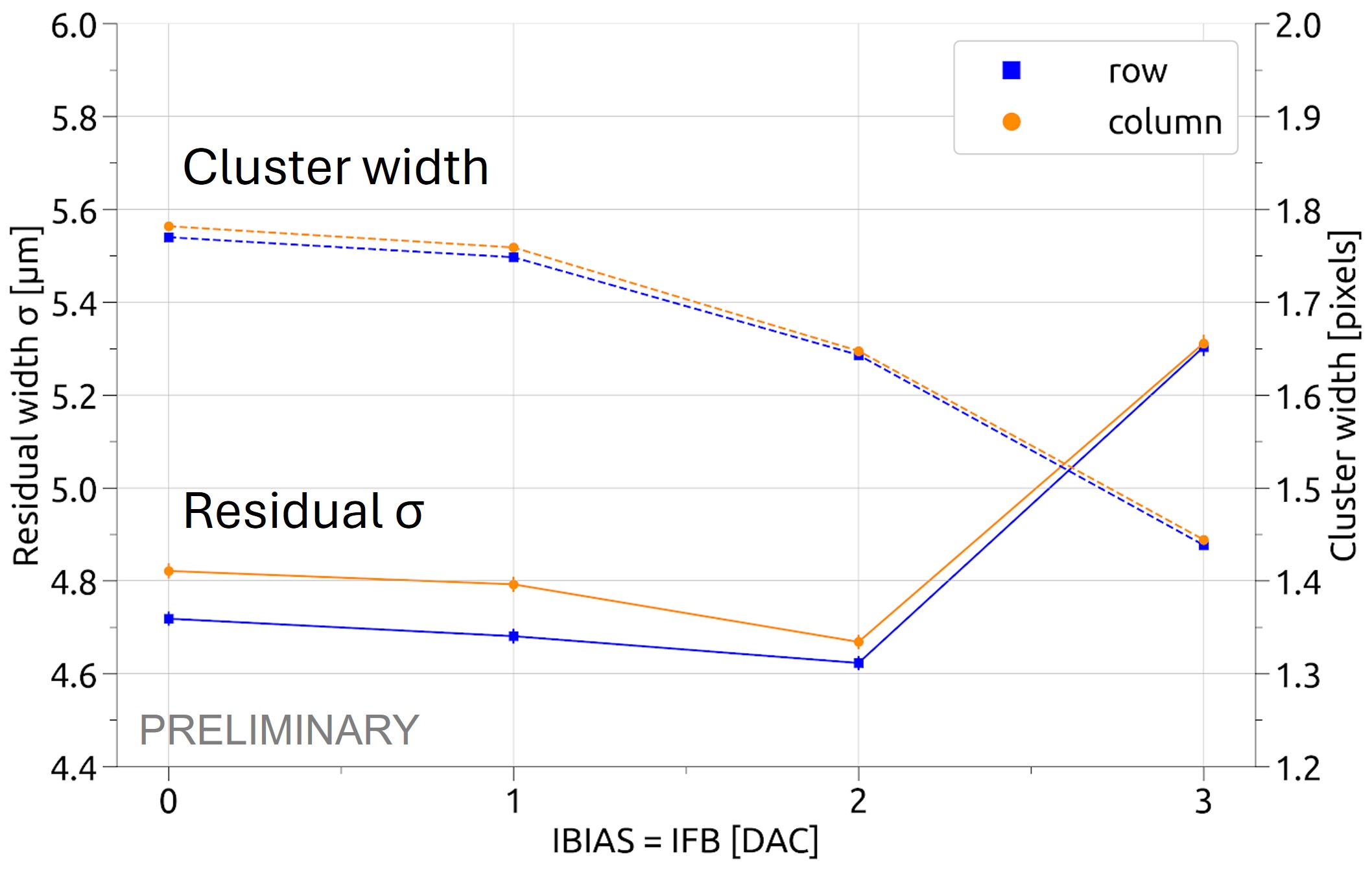}
    \caption{}
    \label{fig:ibias_resolution}
  \end{subfigure}
  \caption{Cluster and residual width along the two directions for ID \subref{fig:id_resolution} and IBIAS-IFB scan \subref{fig:ibias_resolution}}
  \label{fig:FE_resolution}
\end{figure}

\noindent A comparison of the cluster-size distributions is shown in \autoref{fig:clz_id}, where it can be seen that by increasing the ID current, there is a slight increase in the population of single and double-pixel clusters. For \textcolor{black}{the} IBIAS-IFB currents, the same trend is observed with a more prominent increase of single-pixel and double-pixel clusters for IBIAS $=$ IFB $=$ 3, see \autoref{fig:clz_ibias}. \\
In \autoref{fig:FE_resolution}, the average cluster width, i.e. maximum elongation of the cluster, and the residual width, given by the $\upsigma$ of the gaussian fit of the residual distribution, are shown separately for the row and column axis. As shown in \autoref{fig:id_resolution}, the average cluster width slightly decreases by increasing ID. The residual width gets wider; the effect is small since the residual width variation is less than 5\% comparing ID = 0 and ID = 3.\\
As shown in \autoref{fig:ibias_resolution}, the average cluster width decreases significantly \textcolor{black}{when} increasing IBIAS and IFB. The residual width decreases up to IBIAS = IFB = 2. A significant widening of the residual width is visible at IBIAS = IFB = 3. \\
In both scans, the cluster and the residual width are slightly larger on the column with respect to the row axis. This asymmetry is thought to be due to \textcolor{black}{a} small misalignment of the DUT: along the two axes, the residual width does not differ \textcolor{black}{by} more than a few percent. \\
From both scans, the minimum residual width is much smaller than the binary resolution (\SI{7.2}{\micro\meter}) because the average cluster width is always greater than 1. At a cluster width of about 1.65 pixels, the minimum residual width results in a value between 4.6 and 4.7 \si{\micro\meter}.
\section{Conclusions and outlook}
The first in-beam characterization of the ARCADIA MD3, with \SI{200}{\micro\meter} active thickness, was performed at FTBF, leading to early studies of the MD3 tracking performance. The impact of the front-end currents (ID, IBIAS, IFB) \textcolor{black}{was} investigated \textcolor{black}{in terms of cluster and residual width}. \textcolor{black}{It is observed that IBIAS and IFB have a greater impact on the spatial resolution with respect to ID}. \textcolor{black}{Overall,} the minimum width of the residual distribution is \SI{4.6}-\SI{4.7}{\micro\meter} with an average cluster width of 1.65 pixels. \textcolor{black}{The large active thickness of the MD3 enhances the charge sharing and this is essential to improve the spatial resolution with respect to the binary one, which is only determined by the pixel pitch. In terms of front-end optimization, this work shows that also these bias currents do impact on the physical performance of the sensor.}
The analysis aimed at a detailed and complete characterization of the tracking capabilities of the MD3, in terms of efficiency and spatial resolution, is ongoing and further results will be presented.
\acknowledgments
The authors gratefully acknowledge Fermilab accelerator and Test Beam Facility staff for the accelerator performance and the continuous support. \\
The authors declare that financial support was received for the research and/or publication of this article. The authors acknowledge the support provided by INFN. This work was produced by FermiForward Discovery Group, LLC under Contract No. 89243024CSC000002 with the U.S. Department of Energy, Office of Science, Office of High Energy Physics. Publisher acknowledges the U.S. Government license to provide public access under the \href{https://www.energy.gov/downloads/doe-public-access-plan}{DOE Public Access Plan}. \textcolor{black}{This study was carried out within the Space It Up and received funding from the ASI and the MUR – Contract n. 2024-5-E.0 - CUP n. I53D24000060005. This work was supported by the Science Committee of Republic of Armenia (Research projects No.22AA-1C009 and 22rl-037).}

\bibliographystyle{JHEP}
\bibliography{biblio}

\end{document}